\documentclass[12pt]{iopart}
\usepackage{iopams}
\usepackage{graphicx}

\begin{document}

\title{Comparison of work fluctuation relations}

\author{Jordan Horowitz$^1$ and Christopher Jarzynski$^2$}

\address{$^1$ Department of Physics, University of Maryland, College Park, MD 20742 USA}
\address{$^2$ Department of Chemistry and Biochemistry, and Institute for Physical Science and Technology, University of Maryland, College Park, MD 20742 USA}
\ead{horowitz@umd.edu and cjarzyns@umd.edu}

\begin{abstract}
We compare two predictions regarding the microscopic fluctuations of a system
that is driven away from equilibrium:
one due to Crooks [{\it J.\ Stat.\ Phys.} {\bf 90}, 1481 (1998)]
which has gained recent attention in the context of nonequilibrium work and fluctuation
theorems,
and an earlier, analogous result obtained by Bochkov and Kuzovlev [{\it Zh. Eksp. Teor. Fiz.} {\bf 72(1)}, 238Ð247 (1977)].
Both results quantify irreversible behavior
by comparing probabilities of observing particular microscopic trajectories during thermodynamic
processes related by time-reversal, and both
are expressed in terms of the {\it work} performed when driving the system away from equilibrium.
By deriving these two predictions within a single, Hamiltonian framework,
we clarify the precise relationship between them,
and discuss how the different definitions of work used by the two sets of authors
gives rise to different physical interpretations.
We then obtain a extended fluctuation relation that contains both the Crooks
and the Bochkov-Kuzovlev results as special cases.
\end{abstract}

\maketitle

\section{Introduction}

Recent interest in the nonequilibrium thermodynamics of small systems~\cite{Bustamante2005}
has brought attention to a set of papers written
by Bochkov and Kuzovlev in the late 1970's and early
1980's~\cite{Bochkov1977a,Bochkov1977b,Bochkov1981a,Bochkov1981b}.
In view of evident similarities between results derived in these papers,
and predictions obtained over the past decade or so by other authors (see in particular \cite{Jarzynski1997a,Jarzynski1997b,Crooks1998,Crooks1999a,Hummer2001a}),
it is desirable to clarify the precise relationships between between these two sets of results.
In an earlier paper \cite{Jarzynski2007}, one of us has compared the nonequilibrium work relation of \cite{Jarzynski1997a},
$\left\langle \rme^{-\beta W} \right\rangle = \rme^{-\beta\Delta F}$,
with an analogous equality found in \cite{Bochkov1977a,Bochkov1977b,Bochkov1981a,Bochkov1981b},
$\left\langle \rme^{-\beta W_0} \right\rangle = 1$. 
In the present paper we turn our attention to {\it fluctuation relations} for systems driven away from equilibrium. Specifically, we compare a result due to Crooks \cite{Crooks1998} (\eref{eq:intro.crooks} below) with a similar result obtained by Bochkov and Kuzovlev \eref{eq:intro.bk}.

Briefly, the setting is a system driven away from equilibrium by an externally applied, time-dependent force, $X_{t}$.
We consider two different protocols (or schedules) for applying this force,
$X_t^F$ and $X_t^R$, related by time-reversal \eref{eq:conjugateProtocols},
where the superscripts denote ``forward'' and ``reverse''.
The microscopic evolution of the system in either case is described by a phase space trajectory, $\gamma$.
Under appropriate assumptions regarding this evolution,
Crooks has shown that the probability to observe a given trajectory $\gamma^F$ during the forward process,
and the probability to observe its time-reversed counterpart $\gamma^R$ during the reverse process,
are related by~\cite{Crooks1998}
\begin{equation}
\label{eq:intro.crooks}
\frac{P_F[\gamma^{F}]}{P_R[\gamma^{R}]}=\rme^{\beta(W-\Delta F)} .
\end{equation}
Here, $W$ denotes the work performed on the system, and $\Delta F$ is a free energy difference between the equilibrium ensembles from which initial conditions are sampled, as described in detail below.
From this result others are easily derived, including a fluctuation theorem
for the ratio of forward and reverse probability distributions of work values,
$\rho_F(W)$ and $\rho_R(W)$~\cite{Crooks1999a},
which has recently been confirmed experimentally~\cite{Collin2005}.

In a similar context, Bochkov and Kuzovlev have made an analogous assertion
(Equation 2.11 of \cite{Bochkov1981a} or equation 7 of \cite{Bochkov1977b}), which in the notation of the present paper reads,
\begin{equation}
\label{eq:intro.bk}
\frac{P_F^0[\gamma^{F}]}{P_R^0[\gamma^{R}]}=\rme^{\beta W_0} ,
\end{equation}
where $W_0$ again denotes work.

As the above notation suggests, the two definitions of work, $W$ and $W_0$, differ, as do the probability distributions $P_{F/R}$ and $P_{F/R}^0$.
Moreover, $\Delta F$ appears in \eref{eq:intro.crooks},
but not in \eref{eq:intro.bk}.
To establish the relationship between these two predictions, we first derive them both
within a Hamiltonian framework, in \sref{sec:review}.
Following that, in \sref{sec:extension} we derive a more general result,
containing both \eref{eq:intro.crooks} and \eref{eq:intro.bk} as special cases.
This extended relation is obtained for both Hamiltonian dynamics and Markovian stochastic dynamics.

Independently, Seifert~\cite{Seifert2007} has concluded that both \eref{eq:intro.crooks} and \eref{eq:intro.bk}
are special cases of a more general result, obtained within a theoretical framework
that defines entropy production in out-of-equilibrium systems~\cite{Seifert2005b}.
Our approach complements Seifert's, by focusing on the different physical interpretations
of work appearing in \eref{eq:intro.crooks} and \eref{eq:intro.bk}.

\section{Derivations of Earlier Results}\label{sec:review}

In this section we derive \eref{eq:intro.crooks} and \eref{eq:intro.bk} within a single framework,
using Hamilton's equations to model the microscopic evolution of the system of interest.
We introduce and specify the notation for this framework in \sref{sec:setup},
and then in \sref{sec:crooks} and \ref{sec:bk} we obtain the desired results.
These side-by-side derivations clarify the physical interpretation of the differences
between \eref{eq:intro.crooks} and \eref{eq:intro.bk},
as highlighted in a brief summary at the end of this section.
We note that Crooks originally obtained \eref{eq:intro.crooks} by modeling the evolution
of the system as a stochastic, Markov process.
The Hamiltonian derivation of this result presented here is similar to the analyses of
\cite{Cleuren2006a,Jarzynski2006a}.

\subsection{Setup}\label{sec:setup}

Consider a classical system with $N$ degrees of freedom,
described by coordinates $\textbf{q}=(q_{1},\dots,q_{N})$ and conjugate momenta $\textbf{p}=(p_{1},\dots,p_{N}$), and let $z=(\textbf{q},\textbf{p})$ denote a point in its phase space.
We will be interested in the evolution of this system, in the presence
of an externally controlled, time-dependent force $X_t$.
In this section we model this evolution using Hamilton's equations,
assuming a time-dependent Hamiltonian of the form
\begin{equation}
\label{eq:rev:ham}
H(z;X_{t})=H_{0}(z)-X_{t}\alpha(z).
\end{equation}
Here $\alpha$ is the coordinate conjugate to the external force $X$,
and the ``bare'' Hamiltonian $H_0$ denotes the energy of the system in the absence of this force.
For simplicity, we further assume that $H$ is time-reversal-invariant
for any fixed value of the external force, that is,
\begin{equation}
\label{eq:tri}
H(z^*;X) = H(z;X),
\end{equation}
where the asterisk denotes a reversal of momenta, ${\bf p} \rightarrow -{\bf p}$.

We have assumed that $X$ appears {\it linearly}
in the definition of the Hamiltonian, $H = H_0 - X\alpha$, as in the papers by
Bochkov and Kuzovlev~\cite{Bochkov1977a,Bochkov1977b,Bochkov1981a,Bochkov1981b}.
As mentioned by those authors, this assumption can be relaxed to allow for nonlinear coupling $H(z;X)$,
as in \cite{Jarzynski1997a,Jarzynski1997b,Crooks1998,Crooks1999a,Hummer2001a}.
However, for clarity of exposition, we follow 
\cite{Bochkov1977a,Bochkov1977b,Bochkov1981a,Bochkov1981b}
and stay with the linear assumption throughout this paper.

We will use the term {\it process} to denote the following sequence of events.
First, the system is prepared in a state of thermal equilibrium at inverse temperature $\beta$.
(This step will be discussed in more detail below.)
Subsequently, from $t=0$ to $t=\tau$ the system evolves in time under Hamilton's equations,
as the external force is applied according to a schedule, or {\it protocol}, $X_t$.
We will use the notation $z_t$ to specify the phase space coordinates of the system
at a particular time $t$ during this interval, and $\gamma$ to denote the entire trajectory
from $t=0$ to $t=\tau$.
Such a trajectory represents a single {\it realization}
-- i.e.\ one possible microscopic history -- of the process in question.
By performing this process repeatedly -- always preparing the system in equilibrium,
and always applying the same protocol $X_t$ --
we effectively generate independent samples $\gamma_1,\,\gamma_2,\cdots$
from a statistical ensemble of possible realizations of the process.

\begin{figure}[tbp]
\includegraphics[scale=0.5, angle=-90]{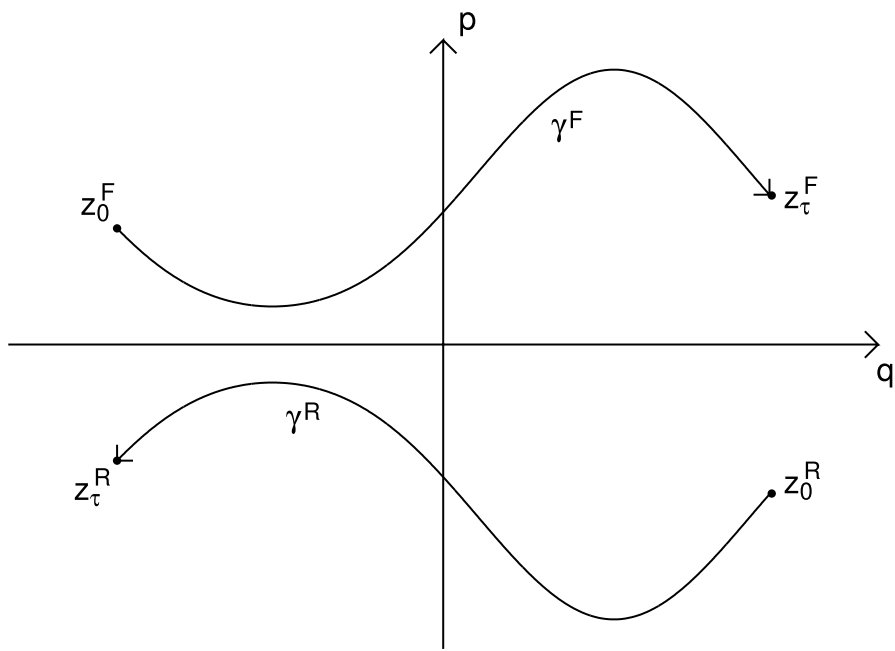}
\centering
\caption{Schematic depiction of a forward trajectory $\gamma^{F}$
and its conjugate twin $\gamma^{R}$.
The two are related by a reversal of momenta and the direction of time,
i.e.\ $z_t^R = z_{\tau-t}^{F*}$ for $0\le t\le\tau$.
}
\label{fig:rev:conjugatepair}
\end{figure}

We will consider two different processes,
which we label {\it forward} ($F$) and {\it reverse} ($R$).
During the forward process the force is switched from
$X_0^F=A$ to $X_\tau^F=B$ using the forward protocol $X_{t}^{F}$,
while during the reverse process the force is switched from $B$ back to $A$,
using the time-reversed protocol, i.e.\
\begin{equation}
\label{eq:conjugateProtocols}
X_t^R = X_{\tau-t}^F.
\end{equation}
Our assumption of time-reversal invariance then implies that solutions of Hamilton's
equations come in {\it conjugate pairs}~\cite{Cleuren2006a,Jarzynski2006a}:
for every trajectory $\gamma^F = \{ z_t^F \}$ that represents a microscopic
realization of the forward process,
the conjugate twin $\gamma^R = \{ z_t^R\}$
is a realization of the reverse process, where
\begin{equation}
\label{eq:conjPair}
z_t^R = z_{\tau-t}^{F*},
\end{equation}
as illustrated in figure \ref{fig:rev:conjugatepair}.
While we restrict our analysis to Hamiltonian dynamics,
we note that conjugate pairing of trajectories holds also for non-Hamiltonian,
time-reversible deterministic dynamics,
as exploited by Evans to derive nonequilibrium work relations
for deterministically thermostatted systems~\cite{Evans2003}.

In the derivations that follow, we will make use of two different definitions
of the {\it work} performed on the system during such a process.
The first definition, which appears in the papers by Bochkov and Kuzovlev
(e.g.\  equation 2.9 of \cite{Bochkov1981a}),
is the integral of force versus displacement
familiar from introductory mechanics texts~\cite{Halliday2005}:
\begin{equation}
\label{eq:bk:work}
W_{0} \equiv
-\int_0^\tau \dot\alpha\,\frac{\partial H}{\partial\alpha}\, {\rm d}t =
\int_{0}^{\tau}X\dot{\alpha}\,{\rm d}t=H_{0}(z_{\tau})-H_{0}(z_{0}).
\end{equation}
By contrast, recent papers (see e.g.\ \cite{Jarzynski1997a,Crooks1998})
have used a different definition,
\begin{equation}
\label{eq:crooks:work}
W \equiv
\int_{0}^{\tau}\dot{X}\frac{\partial H}{\partial X}\,{\rm d}t
=-\int_{0}^{\tau}\dot{X}\alpha \, {\rm d}t=H(z_{\tau}; X_{\tau})-H(z_{0};X_{0})
\end{equation}
which traces its origins to discussions of the statistical foundations
of thermodynamics~\cite{Gibbs1902,Schrodinger1962,Uhlenbeck1963}.
In both \eref{eq:bk:work} and \eref{eq:crooks:work} the properties
of Hamilton's equations have been invoked to rewrite the integral as a net change in
the value of a Hamiltonian, either $H_0$ or $H$(see \cite{Jarzynski2007}).
At the end of this section we briefly discuss the distinction between these
two definitions of work.

\subsection{Crooks Fluctuation Relation}\label{sec:crooks}

For the situation that we have just described, we now derive \eref{eq:intro.crooks}.
In this equation, $P_F[\gamma^F]$ denotes the probability distribution to observe
a trajectory $\gamma^F$, when performing the forward process.
This distribution must be defined with respect to a {\it measure} on the space
of trajectories $\gamma^F$.
Since Hamiltonian evolution is deterministic, it is natural to use the Liouville phase space
measure, applied to the initial conditions of the trajectory.
Thus, if we imagine a narrow ``tube'' of Hamiltonian trajectories evolving
from $t=0$ to $t=\tau$,
from a small patch of initial conditions in phase space,
then we equate the path-space measure of this tube of trajectories
with the Liouville measure of the patch of initial conditions:
${\rm d}\gamma^F = {\rm d}z_0^F = {\rm d}^N\!{\bf q}_0\,{\rm d}^N\!{\bf p}_0$.
Similarly, for the reverse process the distribution $P_R[\gamma^R]$
is defined with respect to the measure ${\rm d}\gamma^R = {\rm d}z_0^R$.
Because Hamiltonian evolution preserves phase space volume~\cite{Goldstein1980},
the measure of a given set of forward trajectories is the same as that
of the conjugate set of reverse trajectories:
\begin{equation}
{\rm d}\gamma^F = {\rm d}\gamma^R.
\end{equation}

In Crooks's formulation~\cite{Crooks1998}, the initial conditions for the forward process
are sampled from the equilibrium (canonical) distribution corresponding to the initial
value of the external force, $X_0^F = A$:
\begin{equation}
\label{eq:PF}
P_F[\gamma^F] = p_A^{\rm eq}(z_0^F) =
\frac{1}{Z(A)}
\exp[-\beta H(z_0^F;A)] .
\end{equation}
Similarly for the reverse process,
\begin{equation}
\label{eq:PR}
P_R[\gamma^R] = p_B^{\rm eq}(z_0^R) =
\frac{1}{Z(B)}
\exp[-\beta H(z_0^R;B)] .
\end{equation}
In these equations
\begin{equation}
\label{eq:crooks:partition}
Z(X) = \int {\rm d}z \, \rme^{-\beta H(z;X)}
\end{equation}
is the classical partition function associated with the canonical distribution
at a fixed value of $X$;
the corresponding free energy is given by the expression
\begin{equation}
\label{eq:FX}
F(X) = -\beta^{-1} \ln Z(X).
\end{equation}
Combining \eref{eq:PF}, \eref{eq:PR}, and \eref{eq:FX}
we get, for a conjugate pair of trajectories $\gamma^F$ and $\gamma^R$,
\begin{equation}
\frac{P_F[\gamma^F]}{P_R[\gamma^R]} =
\rme^{-\beta\Delta F}
\rme^{\beta[ H(z_0^R;B) - H(z_0^F;A) ]},
\end{equation}
where $\Delta F \equiv F(B) - F(A)$.
We now use \eref{eq:tri}, \eref{eq:conjPair}, and \eref{eq:crooks:work}
to rewrite the quantity inside the last exponent above:
\begin{eqnarray}
\label{eq:rewrite}
\nonumber
H(z_0^R;B) - H(z_0^F;A) 
&=& H(z_\tau^{F*};B) - H(z_0^F;A)\\
\nonumber
&=& H(z_\tau^F;B) - H(z_0^F;A) \\
&=& W,
\end{eqnarray}
where $W$ represents the work (as defined by \eref{eq:crooks:work})
performed on the system during the realization $\gamma^F$.
Thus we finally obtain
\begin{equation}
\label{eq:crooks}
\frac{P_F[\gamma^F]}{P_R[\gamma^R]} =
\rme^{\beta(W-\Delta F)} .
\end{equation}
This result was first derived by Crooks~\cite{Crooks1998}, who modeled the evolution
of the system as a stochastic discrete time Markov process, rather than using Hamiltonian dynamics.
One consequence of \eref{eq:crooks} is the Crooks Fluctuation Theorem for the ratio of forward and reverse work distributions~\cite{Crooks1999a}:
\begin{equation}
\label{eq:cft}
\frac{\rho_{F}(W)}{\rho_{R}(-W)}=\rme^{\beta(W-\Delta F)}.
\end{equation}
Often \eref{eq:crooks} and \eref{eq:cft} are expressed in terms of the dissipated work, $W_{\rm d}=W-\Delta F$~\cite{Crooks2000}.

\subsection{Bochkov-Kuzovlev Fluctuation Relation}\label{sec:bk}

In deriving \eref{eq:crooks}, we imagined that initial conditions
were sampled from the canonical distributions
associated with the Hamiltonians $H(z;A)$ and $H(z;B)$,
for the forward and reverse processes, respectively.
Now let us consider a different situation, in which -- for both processes --
initial conditions are sampled from the canonical distributions
corresponding to the bare Hamiltonian $H_0(z)$.
We suppose, however, that the protocols $X_t^F$ and $X_t^R$ are the same
as above, namely from $A$ to $B$ and from $B$ to $A$, where
these values remain arbitrary.
If we use $P_F^0[\gamma^F]$ and $P_R^0[\gamma^R]$ to denote the corresponding
distributions of trajectories, then we have
\begin{eqnarray}
\label{eq:traj_prob1}
P_F^0[\gamma^F] =
\frac{1}{Z(0)}
\exp[-\beta H_0(z_0^F)],\\
\label{eq:traj_prob2}
P_R^0[\gamma^R] =
\frac{1}{Z(0)}
\exp[-\beta H_0(z_0^R)]
\end{eqnarray}
in place of \eref{eq:PF}, \eref{eq:PR}.
Taking the ratio then gives us
\begin{equation}
\frac{P_F^0[\gamma^F]}{P_R^0[\gamma^R]} =
\rme^{\beta[ H_0(z_0^R) - H_0(z_0^F) ]}.
\end{equation}
By analogy with \eref{eq:rewrite}, we have
\begin{equation}
H_0(z_0^R) - H_0(z_0^F)
= H_0(z_\tau^F) - H_0(z_0^F)
= W_0,
\end{equation}
where $W_0$ is the work (as defined by \eref{eq:bk:work})
performed during the realization $\gamma^F$.
Thus we finally arrive at
\begin{equation}
\label{eq:bk}
\frac{P_F^0[\gamma^F]}{P_R^0[\gamma^R]} =
e^{\beta W_0},
\end{equation}
which corresponds to equation 2.11 of \cite{Bochkov1981a}.

Related to \eref{eq:bk} is a fluctuation theorem for the work distributions (cf. \eref{eq:cft}),
\begin{equation}
\label{eq:bk_cft}
\frac{\rho_{F}^{0}(W_{0})}{\rho_{R}^{0}(-W_{0})}=\rme^{\beta W_{0}},
\end{equation}
where $\rho_{F/R}^{0}(W_{0})$ are the work distributions associated with the trajectory distributions in \eref{eq:traj_prob1} and \eref{eq:traj_prob2}.  Note that the quantity $W_{0}$ that appears in \eref{eq:bk} and \eref{eq:bk_cft} is in general different from the dissipated work $W_{\rm d}$.

\vskip .2in

As highlighted by these derivations, a principal
difference between \eref{eq:crooks} and \eref{eq:bk}
(equivalently, \eref{eq:intro.crooks} and \eref{eq:intro.bk})
is the choice of the equilibrium distributions from which initial conditions are drawn.
If we sample from canonical distributions corresponding
to the Hamiltonians $H(A)$ and $H(B)$, then we obtain \eref{eq:crooks},
while if we sample initial conditions from the bare canonical distribution,
corresponding to $H_0$, then
we arrive at \eref{eq:bk}.

It is useful to spell out the physical interpretation associated with this distinction.
In Crooks's formulation \eref{eq:crooks},
we imagine that prior to $t=0$ we allow the system to come to
equilibrium with a thermal reservoir at inverse temperature $\beta$,
while holding the external force fixed at either $A$ or $B$.
We then let the system evolve
as the external force is manipulated according to the forward or reverse protocol.
To obtain Bochkov and Kuzovlev's result \eref{eq:bk},
we instead imagine that the external force
is absent ($X=0$) as the system equilibrates with the reservoir.
The system thus settles into an equilibrium state corresponding to the bare
Hamiltonian $H_0$.
Just prior to $t=0$ we suddenly turn the force on to $A$ or $B$,
subsequently switching it continuously according to the forward or reverse protocol.

The two results involve not only different distributions of initial conditions,
but also different definitions of work, $W_0$ and $W$.
The right sides of \eref{eq:bk:work} and \eref{eq:crooks:work} suggest
a physical interpretation of this distinction.
Namely, in one case we define the {\it internal energy} of the system
to be given by the value of the bare Hamiltonian $H_0$;
then $W_0$ is viewed as the work performed by the application of an external force
$X$ that affects the evolution of the system in a fixed energy landscape $H_0$.
In the other case, the system's internal energy is taken to be the full
Hamiltonian $H$;
we then view $W$ as the work associated with externally controlled manipulations
of the energy landscape $H=H_0-X\alpha$.
See \cite{Jarzynski2007}
for a more detailed discussion, and illustration, of this distinction.

In the special case of cyclic processes, defined by the condition $X_{0}=X_{\tau}$, we have $\Delta F=0$ and $W=W_{0}$~\cite{Jarzynski2007}.  
In this situation \eref{eq:crooks} and \eref{eq:bk} are equivalent.
If we  further restrict ourselves to time symmetric protocols ($X_{t}=X_{\tau-t}$), then there is no distinction between the forward and reverse processes.  In this case, \eref{eq:crooks} and \eref{eq:bk} can be viewed as special cases of a more general result derived by Evans and Searles (see equation (4.15) of \cite{Evans2002a}, as well as the recent review by Sevick {\it et al} \cite{Sevick2007}).

\section{Extended Relation}\label{sec:extension}

In \sref{sec:review} we used Hamiltonian dynamics to model the evolution
of the system.
By contrast, as noted above, Crooks modeled this evolution as a
discrete-time, stochastic Markov process, to represent a system in contact with
a thermal reservoir~\cite{Crooks1998}.
If we were to carry out an analysis similar to that of \sref{sec:review},
but in a Markovian rather than a Hamiltonian framework,
then we would arrive at the same conclusion, namely that
both \eref{eq:intro.crooks} and \eref{eq:intro.bk} are valid,
the difference between them relating to the choice of canonical ensemble
from which to sample initial conditions.
Rather than showing this explicitly, in this section we derive a more general result,
\eref{eq:combo} below,
that contains both \eref{eq:intro.crooks} and \eref{eq:intro.bk} as special cases.
We will present both Hamiltonian and Markovian derivations of this result.

Consider a Hamiltonian of the form
\begin{equation}
\label{eq:ext:hamiltonian}
H(z;\lambda_{t},X_{t})=H_{0}(z;\lambda_{t})-X_{t}\alpha(z),
\end{equation}
where now $H_0$ itself is parametrized by a variable $\lambda$.
Thus the full Hamiltonian $H$ is made time-dependent by manipulating
both the force $X$, and a {\it work parameter}, $\lambda$.
From a mathematical perspective, this division of the parameters into an external force
and an internal work parameter is somewhat arbitrary,
particularly if $\lambda$ appears linearly in $H_0$ (as in \eref{eq:discH} below).
While in specific situations the distinction may be motivated by physical considerations or by the questions being addressed, our aim here is to develop a general formalism that encompasses both \eref{eq:intro.crooks} and \eref{eq:intro.bk}.
For this purpose, the form of $H$ given in \eref{eq:ext:hamiltonian} is useful.

For the bare Hamiltonian $H_0$, we have a parameter-dependent canonical distribution
\begin{equation}
\label{eq:p0}
\tilde{p}_0^{\rm eq}(z;\lambda) =
\frac{1}{\tilde Z(\lambda)} \, \rme^{-\beta H_{0}(z;\lambda)} ,
\end{equation} 
for which the partition function and free energy are given by
\begin{equation}
\label{eq:ext:partition}
\tilde Z(\lambda) = \int {\rm d}z \, \rme^{-\beta H_{0}(z;\lambda)}
\qquad\textrm{and}\qquad
\tilde F(\lambda)=-\beta^{-1}\ln \tilde Z(\lambda).
\end{equation}

Throughout this section we will take the internal energy of the system to be defined
by the value of the bare Hamiltonian $H_0$, and
we will consider processes during which the system evolves in time
as both $\lambda$ and $X$ are switched according to pre-determined protocols.
Thus we picture a system that evolves in a time-dependent energy landscape $H_0$,
while also coupled to a time-dependent external force $X$.
For a given realization of such a process, we take the work performed on the system
to be given by
\begin{equation}
\label{eq:ext:work}
\tilde{W}=\int_{0}^{\tau}\dot{\lambda}\frac{\partial H}{\partial \lambda} \, {\rm d}t
+\int_{0}^{\tau}X\dot{\alpha} \, {\rm d}t,
\end{equation}
or by the discrete-time analogue of this expression,
\eref{eq:discreteWork} below.
This is a hybrid definition whose two terms are similar to the expressions
for $W$ and $W_0$ introduced earlier.
As we will see,
for processes during which the system is thermodynamically isolated
the value of $\tilde W$ is equal to the net change in $H_0$, \eref{eq:wdh0};
whereas if the system remains in contact with a reservoir as $\lambda$
and $X$ are varied,
then the net change in $H_0$ is equal to the sum of $\tilde W$ and a quantity $Q$
that represents the heat absorbed by the system, \eref{eq:deltaH0}.
In either case, \eref{eq:ext:work}
provides a definition of work that is faithful to 
the first law of thermodynamics, and consistent with
our definition of internal energy, $H_0$.

As before, we will compare two processes ($F$, $R$),
characterized by conjugate protocols for the externally controlled parameters:
\begin{equation}
X_t^R = X_{\tau-t}^F
\qquad,\qquad
\lambda_t^R = \lambda_{\tau-t}^F.
\end{equation}
For the forward process, we let $A$ and $B$ denote the initial and final values of $X$, as before,
and we let $0$ and $1$ denote the initial and final values of $\lambda$.
For the reverse process, then,
$X$ is switched from $B$ to $A$, and $\lambda$ from 1 to 0.
We will assume that initial conditions for the forward and reverse processes
are sampled from
$p_0^{\rm eq}(z;\lambda=0)$ and $p_0^{\rm eq}(z;\lambda=1)$, respectively
(see \eref{eq:p0}).
The physical interpretation of this assumption is similar
to that discussed in the context of the Bochkov-Kuzovlev results, near the end of \sref{sec:review}.
Namely, we imagine that the external force is absent ($X=0$) as the system equilibrates with the reservoir prior to time $t=0$.
The work parameter is held fixed at either $\lambda=0$ (for the forward process) or $\lambda=1$ (for the reverse process) during this equilibration stage.
Then, immediately prior to $t=0$ the external force is turned on to either $A$ or $B$, and subsequently both $\lambda$ and $X$ are varied according to the appropriate protocol.

\subsection{Hamiltonian dynamics} \label{sec:hamiltonian}

With the setup just described, we now suppose that the microscopic history of the system
during the time interval $0 \le t \le\tau$ (for either process) is described by a Hamiltonian trajectory
$z_t$.
Such evolution has the property that the total rate of change of the Hamiltonian is
equal to the partial derivative of the Hamiltonian function with respect to time
(see e.g.\ equations 8-35 of \cite{Goldstein1980}).
Thus
\begin{eqnarray}
\frac{{\rm d}}{{\rm d}t} H(z_t;\lambda_t,X_t) &=&
\dot\lambda_t \frac{\partial H}{\partial\lambda} + 
\dot X_t \frac{\partial H}{\partial X} \\
&=&
\dot\lambda_t \frac{\partial H_0}{\partial\lambda} - \dot X \alpha \\
&=&
\dot\lambda_t \frac{\partial H_0}{\partial\lambda} + X \dot\alpha
- \frac{{\rm d}}{{\rm d}t} \left(X\alpha\right) .
\end{eqnarray}
Adding $({\rm d}/{\rm d}t) (X\alpha)$ to both sides, we get
\begin{equation}
\label{eq:dh0dt}
\frac{{\rm d}}{{\rm d}t} H_0 = \dot\lambda_t \frac{\partial H_0}{\partial\lambda} + X \dot\alpha.
\end{equation}
The first two terms on the right are the integrands appearing in
our definition of $\tilde W$ \eref{eq:ext:work}.
Thus, integrating both sides of \eref{eq:dh0dt} with respect to time,
we see that $\tilde W$ is the net change in the value of the bare Hamiltonian:
\begin{equation}
\label{eq:wdh0}
\tilde W = H_{0}(z_{\tau};\lambda_{\tau})-H_{0}(z_{0};\lambda_{0}).
\end{equation}
This identity will now be put to use much as \eref{eq:bk:work}
and \eref{eq:crooks:work} were in \sref{sec:review}.

Adopting notation similar to that of \sref{sec:review}, we
let $\tilde{P}_F[\gamma^F]$ and $\tilde{P}_R[\gamma^R]$ denote the probability distributions
of realizations of the forward and reverse processes;
as before we use the Liouville measure of initial conditions to define
a measure on the space of possible trajectories, for both processes.
We then have, by analogy with \eref{eq:PF} and \eref{eq:PR},
\begin{eqnarray}
\label{eq:ext:PF}
\tilde{P}_F[\gamma^F] = \tilde{p}_0^{\rm eq}(z_0^F;0) =
\frac{1}{\tilde Z(0)}
\exp[-\beta H_0(z_0^F;0)]
\\
\label{eq:ext:PR}
\tilde{P}_R[\gamma^R] = \tilde{p}_0^{\rm eq}(z_0^R;1) =
\frac{1}{\tilde Z(1)}
\exp[-\beta H_0(z_0^R;1)] ,
\end{eqnarray}
since initial conditions are sampled from the canonical
ensembles associated with $H_0$.
Taking the ratio of these two distributions for a conjugate pair of trajectories,
and making use of \eref{eq:ext:partition} and \eref{eq:wdh0},
we obtain
\begin{equation}
\label{eq:combo}
\frac{\tilde{P}_F[\gamma^F]}{\tilde{P}_R[\gamma^R]} =
\rme^{\beta(\tilde W - \Delta \tilde F)} ,
\end{equation}
where $\Delta \tilde F = \tilde F(1) - \tilde F(0)$.

\subsection{Markovian Stochastic Dynamics}\label{sec:markovian}

We now derive the same result, \eref{eq:combo}, for a system that remains in weak
contact with a thermal reservoir during the interval $0\le t\le\tau$,
and whose evolution in this situation
is modeled as a discrete-time Markov process.
Specifically, we imagine that the microscopic history of the system is described
by a sequence
$z_0 , z_1, \cdots z_N$,
representing phase space points visited at times
$t_0, t_1, \cdots t_N$, with $t_n = n\tau/N$.
With each time increment, both the phase space point $z$ and the parameters
 $\pi \equiv (\lambda,X)$ are updated with a combination of a stochastic step and a switching step,
as follows.
 
 {\it Stochastic step}.
During a stochastic step, starting from, say, $z_n$ at time $t_n$,
the system makes a random jump to a new point
in phase space, $z_{n+1}$, according to the transition probability
$P(z_n\to z_{n+1} ; \lambda,X)$,
where $(\lambda,X)$ are the fixed values of the external parameters
during this step.
This transition probability is assumed to satisfy detailed balance
(a requirement that may be reduced, at the expense of using
a modified dynamics for the reverse process~\cite{Crooks2000}),
expressed as
\begin{equation}
\label{eq:detailedBalance}
\frac{P(z_{n}\to z_{n+1}; \lambda,X )}{P(z_{n}\gets z_{n+1};\lambda,X)}
=\frac{\rme^{-\beta H(z_{n+1};\lambda,X)}}{\rme^{-\beta H(z_{n};\lambda,X)}}.
\end{equation}
The direction of the arrow indicates the sense of the transition,
e.g.\ from $z_{n+1}$ to $z_{n}$ in the denominator.
Since each stochastic step, $z_n\to z_{n+1}$, is meant to represent effects of the
interaction with the thermal reservoir,
any change in the system's energy during such a step will be interpreted
as heat exchanged with the reservoir; see \eref{eq:heat} below.

Detailed balance \eref{eq:detailedBalance} represents an assumption about the dynamics of the system.
Among other things it implies that the equilibrium distribution is conserved when the parameters $\lambda$ and $X$ are held fixed.
Since the equilibrium distribution is determined by all the physical forces acting on the system,
both those that we choose to view as internal and those that we treat as external,
detailed balance is expressed in terms of the full Hamiltonian, $H$, rather than $H_0$.

{\it Switching step}.
During a switching step the external parameters are updated,
e.g.\ from $\pi_n=(\lambda_n,X_n)$ to $\pi_{n+1}=(\lambda_{n+1},X_{n+1})$,
according to a given protocol, with the phase space point held fixed.

The evolution of the system from $t_0=0$ to $t_N=\tau$ consists of 
an alternating sequence of stochastic and switching steps.
For the forward process we assume that the switching step comes first,
giving us a progression of the form
$(z_n^F,\pi_n^F) \to (z_n^F,\pi_{n+1}^F)\to (z_{n+1}^F,\pi_{n+1}^F)$
at each time step.
For the reverse process, the stochastic steps leads:
$(z_n^R,\pi_n^R) \to (z_{n+1}^R,\pi_n^R)\to (z_{n+1}^R,\pi_{n+1}^R)$.

We now define the heat absorbed by the system from the reservoir,
for a given realization of the forward process,
as a sum of energy changes during the stochastic steps:
\begin{equation}
\label{eq:heat}
Q=\sum_{n=0}^{N-1} \delta Q_n = \sum_{n=0}^{N-1}
\left[H(z_{n+1};\pi_{n+1})-H(z_n;\pi_{n+1})\right].
\end{equation}
For the same realization, the work performed on the system is
defined as the discrete-time version of \eref{eq:ext:work}:
\begin{eqnarray}
\label{eq:discreteWork}
\nonumber
\tilde{W}&=&\sum_{n=0}^{N-1} \delta\tilde W_n \\
&=& \sum_{n=0}^{N-1}
\big[H_{0}(z_{n};\lambda_{n+1})-H_{0}(z_{n};\lambda_{n})\big]+
\sum_{n=0}^{N-1}\big[(\alpha_{n+1}-\alpha_{n})X_{n+1}\big],
\end{eqnarray}
introducing the convenient shorthand $\alpha_n \equiv \alpha(z_n)$.
On the right side of \eref{eq:discreteWork},
the first sum represents energy changes due to the switching of
$\lambda$, while the second sum is associated
with the force $X$ acting on the coordinate $\alpha$.
From these definitions, the identity
\begin{eqnarray}
\label{eq:deltaH}
\nonumber
\delta H_n &\equiv&
H(z_{n+1};\pi_{n+1})-H(z_{n};\pi_{n})\\
&=& \delta Q_n + \delta\tilde W_n
+ \alpha_n X_n - \alpha_{n+1} X_{n+1}
\end{eqnarray}
is easily verified, by inspection.
Combining \eref{eq:heat}, \eref{eq:discreteWork} and \eref{eq:deltaH}
now gives us the following expressions for the net changes in $H$ and $H_0$,
from $t=0$ and $t=\tau$:
\begin{eqnarray}
H(z_{N};\pi_{N})-H(z_{0};\pi_{0})
= Q+\tilde{W}+X_{0}\alpha_0-X_N\alpha_N \\
\label{eq:deltaH0}
H_0(z_{N};\lambda_{N})-H_0(z_{0};\lambda_{0})
= Q + \tilde W.
\end{eqnarray}
The latter is analogous to \eref{eq:wdh0}, but now includes heat exchange with the reservoir.

With these preliminaries out of the way, we analyze the probability distributions
of trajectories, for the forward and reverse processes.
A trajectory is specified as a sequence of $N+1$ points in phase space, e.g.\ 
$\gamma = (z_0,z_1, \cdots z_N)$.
A natural measure on the space of such trajectories is given by
${\rm d}\gamma = {\rm d}z_0\,{\rm d}z_1\,\cdots{\rm d}z_N$.
The probability of obtaining a particular trajectory, $\tilde{P}[\gamma]$,
is equal to the probability of sampling its initial conditions $z_0$
from the appropriate equilibrium distribution,
multiplied by the conditional probability of generating the subsequent sequence of steps,
$\tilde{P}[\gamma \vert z_0]$.
Thus for the forward process we have
\begin{eqnarray}
\label{eq:probTraj}
\tilde{P}_F[\gamma^F] = \tilde{p}_0^{\rm eq}(z_0^F;\lambda_0^F=0) \cdot
\tilde{P}_F[\gamma^F \vert z_0^F] \\
\label{eq:condprobTraj}
\tilde{P}_F[\gamma^F \vert z_0^F] =
P(z_0^F \rightarrow z_1^F ; \pi_1^F ) \cdots
P(z_{N-1}^F \rightarrow z_N^F ; \pi_N^F )
\end{eqnarray}
For the reverse process we have analogous equations, except that $\lambda_0^R=1$,
and the transition probabilities are of the form
$P(z_n^R \rightarrow z_{n+1}^R ; \pi_n^R )$,
reflecting the convention that the stochastic step precedes the switching step,
during each time interval of the reverse process.
In writing the conditional probability as the product of transition rates in \eref{eq:condprobTraj},
we have made use of the assumed Markov nature of the dynamics.

For a conjugate pair of trajectories $\gamma^F$ and $\gamma^R$,
the ratio of conditional probabilities is:
\begin{equation}
\label{eq:ratioCondProb}
\frac{\tilde{P}_F[\gamma^F \vert z_0^F]}{\tilde{P}_R[\gamma^R \vert z_0^R]}
=
\frac{
P(z_0^F \rightarrow z_1^F ; \pi_1^F ) \cdots
P(z_{N-1}^F \rightarrow z_N^F ; \pi_N^F )
}
{
P(z_0^F \leftarrow z_1^F ; \pi_1^F ) \cdots
P(z_{N-1}^F \leftarrow z_N^F ; \pi_N^F )
} ,
\end{equation}
where we have used the relations
$z_n^R = z_{N-n}^F$ and $\pi_n^R = \pi_{N-n}^F$ to rewrite
$P_R[\gamma^R \vert z_0^R]$ in terms of $z_n^F$'s and $\pi_n^F$'s.
Following Crooks~\cite{Crooks1998}, we now
use detailed balance to express the right side of \eref{eq:ratioCondProb} as a product
of factors $\exp(-\beta\delta Q_n)$ (see \eref{eq:detailedBalance}, \eref{eq:heat}),
which gives us
\begin{equation}
\frac{\tilde{P}_F[\gamma^F \vert z_0^F]}{\tilde{P}_R[\gamma^R \vert z_0^R]} = \rme^{-\beta Q}.
\end{equation}
To get the ratio of unconditional probabilities $P_F[\gamma^F]/P_R[\gamma^R]$,
we multiply the numerator and denominator of both sides by the appropriate
initial distributions, as per \eref{eq:probTraj}:
\begin{equation}
\frac{\tilde{P}_F[\gamma^F]}{\tilde{P}_R[\gamma^R]}
=
\frac{\tilde{p}_0^{\rm eq}(z_0^F;0)}{\tilde{p}_0^{\rm eq}(z_0^R;1)}
\cdot \rme^{-\beta Q}
= \rme^{\beta \left[ H_0(z_N^F;1) - H_0(z_0^F;0) - \Delta \tilde F - Q \right]} ,
\end{equation}
using $z_0^R = z_N^F$.
With \eref{eq:deltaH0} this becomes
\begin{equation}
\label{eq:combo_again}
\frac{\tilde{P}_F[\gamma^F]}{\tilde{P}_R[\gamma^R]}
= \rme^{\beta (\tilde W - \Delta \tilde F)},
\end{equation}
as in the Hamiltonian case \eref{eq:combo}.

\subsection{Discussion} \label{sec:discussion}

We now argue that \eref{eq:intro.crooks} and \eref{eq:intro.bk} can be viewed
as special cases of \eref{eq:combo}.
To establish this, let us suppose that the parameter $\lambda$ appears linearly
in $H_0$, so that
the full, time-dependent Hamiltonian takes the form
\begin{equation}
\label{eq:discH}
H(z;\lambda_t,X_t) =
\mathcal{H}_{0}(z) - \lambda_t\,\eta(z) - X_t\,\alpha(z).
\end{equation}
It is natural to view both $\lambda$ and $X$ as forces, coupled to the coordinates
$\eta$ and $\alpha$, respectively.
However, in keeping with the convention adopted throughout \sref{sec:extension},
we treat the term $-\lambda\eta$ as a contribution to the {\it internal} energy of the system,
$H_0 = \mathcal{H}_{0}-\lambda\eta$,
and $X$ as an {\it external} force.
The work, \eref{eq:ext:work}, is now given by
\begin{equation}
\label{eq:discW}
\tilde{W}=-\int_{0}^{\tau}\dot{\lambda}\eta \, {\rm d}t
+\int_{0}^{\tau}X\dot{\alpha} \, {\rm d}t,
\end{equation}
or by the corresponding discrete-time analogue obtained from \eref{eq:discreteWork}.
With these definitions \eref{eq:combo} becomes, explicitly,
\begin{equation}
\label{eq:disc.result}
\frac{\tilde{P}_F[\gamma^F]}{\tilde{P}_R[\gamma^R]}
=
\exp\left[
\beta\left(
-\int_{0}^{\tau}\dot{\lambda}\eta \, {\rm d}t
+\int_{0}^{\tau}X\dot{\alpha} \, {\rm d}t
- \Delta\tilde F
\right)\right] .
\end{equation}
As before, initial conditions are sampled from the canonical ensemble
$\tilde{p}_0^{\rm eq} \propto e^{-\beta H_0}$, associated with the initial value of $\lambda$ ($\lambda_{0}^{F}=0$ and $\lambda_{0}^{R}=1$).

If we now consider the special case in which $\lambda=0$ during the entire process
(hence $\tilde{p}_0^{\rm eq} \propto e^{-\beta \mathcal{H}_{0}}$ and $\Delta\tilde F=0$),
then the right side of \eref{eq:disc.result} becomes
$\exp\left( \beta\int X\dot\alpha\,{\rm d}t \right)$, and we recover \eref{eq:intro.bk}.
If we instead imagine that $X=0$ during the entire process,
then the right side of \eref{eq:disc.result} becomes
$\exp\left[\beta \left( -\int \dot\lambda\eta\,{\rm d}t -\Delta\tilde F\right) \right]$.
We then recover \eref{eq:intro.crooks}, with $\lambda$ and $\eta(z)$
playing the roles assigned to $X$ and $\alpha(z)$ in \sref{sec:crooks}.

To summarize, if we view \eref{eq:intro.crooks} as a result that applies when
a system evolves in a time-dependent energy landscape,
and \eref{eq:intro.bk} as a result that applies when a system evolves
under the influence of a time-dependent external force
(as per the discussion at the end of \sref{sec:review}),
then \eref{eq:combo} is a generalization that applies
when both factors are present.

\section{Conclusion}
\label{sec:conclusion}

Our central aim has been to clarify the relationship between the fluctuation relation
obtained by Crooks, \eref{eq:intro.crooks}, and that due to Bochkov and Kuzovlev, \eref{eq:intro.bk}.
These two results share a similar structure, but are not equivalent.
The distinction between them is traced to the different definitions of work ($W$, $W_0$) used by the authors.
One consequence of this difference is that \eref{eq:intro.crooks} involves
(and can be used to estimate) a free energy difference, $\Delta F$,
while \eref{eq:intro.bk} does not.
Moreover, the two results are associated with different definitions of the internal energy of the system,
and rely on different assumptions regarding the initial equilibrium preparation of the system.
Finally, we have obtained a generalized fluctuation relation \eref{eq:combo}
that unites \eref{eq:intro.crooks} and \eref{eq:intro.bk}.

\ack
We gratefully acknowledge financial support provided by the University of Maryland. 

\section*{References}

\bibliography{FluctuationTheory_plus}
\bibliographystyle{iopart-num}

\end{document}